\pdfoutput=1
\documentclass[a4paper, 10pt, conference]{ieeeconf}  

\IEEEoverridecommandlockouts                              

\usepackage{cite}
\usepackage{mathtools}
\usepackage{amsmath,amssymb,amsfonts}
\usepackage{algorithmic}
\usepackage{graphicx}
\usepackage{textcomp}
\usepackage{xcolor}
\usepackage{algorithmic}
\usepackage[ruled,vlined]{algorithm2e}
\usepackage{xcolor}
\usepackage{hyperref}

\title{\LARGE \bf
Towards Boosting the Channel Attention in Real Image Denoising : Sub-band Pyramid Attention}

\author{Huayu Li, Haiyu Wu, Xiwen Chen, 
        Hanning Zhang, and Abolfazl Razi,~\IEEEmembership{Senior Member,~IEEE}
\thanks{H. Li, X. Cheng, and A. Razi was with School of Informatics, Computing and Cyber Systems, Northern Arizona University, Flagstaff, AZ 86011.}
\thanks{H. Wu and H. Zhang are with Xi'an Network Computing Data Technology Co., Ltd., Xi'an 710049, China.}
\thanks{Corresponding author: Abolfazl Razi (e-mail: abolfazl.razi@nau.edu).}
\thanks{H. Li and H. Wu equally contributed to the paper.}
}

\begin{document}

\maketitle
\thispagestyle{empty}
\pagestyle{empty}

\begin{abstract}

Convolutional layers in Artificial Neural Networks (ANN) treat the channel features equally without feature selection flexibility. While using ANNs for image denoising in real-world applications with unknown noise distributions, particularly structured noise with learnable patterns, modeling informative features can substantially boost the performance. Channel attention methods in real image denoising tasks exploit dependencies between the feature channels, hence being a frequency component filtering mechanism. Existing channel attention modules typically use global statics as descriptors to learn the inter-channel correlations. This method deems inefficient at learning representative coefficients for re-scaling the channels in frequency level. This paper proposes a novel Sub-band Pyramid Attention (SPA) based on wavelet sub-band pyramid to recalibrate the frequency components of the extracted features in a more fine-grained fashion. We equip the SPA blocks on a network designed for real image denoising. Experimental results show that the proposed method achieves a remarkable improvement than the benchmark naive channel attention block. Furthermore, our results show how the pyramid level affects the performance of the SPA blocks and exhibits favorable generalization capability for the SPA blocks. 

\end{abstract}

\section{Introduction}


Convolutional Neural Networks (CNN) have shown a remarkable performance in image denoising tasks compared to conventional filtering methods ~\cite{zhang2017beyond, zhang2017learning, mao2016image, anwar2019real}. Mathematically, image denoising methods are supposed to recover a clean and high-quality target $y$ from a corrupted low-quality observation $x$ by eliminating imaging artifacts, distortions, etc., typically modeled as an additive noise $n$. In general, the noise is modeled as an additive white Gaussian Noise (AWGN) with zero mean and a specific variance, the so-called noise level. Using Gaussian noise model provides computational convenience and relies on the central limit theorem that suggests the normalized sum of independent and arbitrarily distributed random terms approach a normal distribution. Although appropriate for general and pure random noise modeling, the AWGN model can be oversimplified for situations where the structured noise may exhibit hidden patterns. For instance, image distortions due to cameras' loss of focus, lens scratch, dusty lens, camera shake, low illumination, raindrops can exhibit more structured and learnable patterns~\cite{boie1992analysis}. 

Previous CNN-based methods~\cite{zhang2017beyond, zhang2018ffdnet, zhang2017learning} outperform the traditional methods~\cite{dabov2007image, buades2005non} due to the powerful learning ability of CNNs. However, they are mainly designed to deal with synthetic noise instead of real-world noise which has more complicated and diverse compositions due to the environmental factors and the utilized image processing pipeline. Therefore, their performance might be suboptimal when applied to real-world denoising tasks. Recently, new models~\cite{anwar2019real, zhao2019pyramid, bao2020real, zamir2020cycleisp} and new benchmarks~\cite{abdelhamed2018high, plotz2017benchmarking} are proposed to tackle real-world noise. One popular way to implement adaptive deep learning network agriculture to deal with real-world noise is to use attention mechanisms to manage and quantify the interdependence. For instance, in RIDNet~\cite{anwar2019real}, a channel attention block~\cite{hu2018squeeze} called feature attention was used for feature selection, which lets the network focus on the feature channels of interest. Although the feature attention obtained a superior performance on real image denoising tasks, generating coarse feature descriptors based on the global pooling of the entire feature map is not optimal.


From the traditional image denoising perspective, a natural image is composed of different frequency components, where the high-frequency components represent the fine details, and low-frequency components represent the global structures. Most DL-based denoising methods over-estimate the learning ability of CNNs and ignore traditional contributions. Thus, we are trying to re-think the DL-based denoising paradigm by leveraging traditional denoising concepts. Channel attention mechanism can be regarded as an adaptive filter that suppresses the abundant frequency feature channels~\cite{zhang2018image}. In most channel attention models, coarse statistics of the feature maps are generated by global average pooling (GAP) as the representatives of the information contained in each feature map. Second-order channel attention was proposed in~\cite{dai2019second} for enriching the representational ability of the channel attention blocks. Nevertheless, these types of channel attention models are not flexible enough to deal with various frequency levels. In ~\cite{anwar2020densely}, Laplacian Attention is proposed as a representative method to integrate traditional methods and deep learning, adopting multiple convolutional layers with different receptive fields to model the frequency components of the input features. Despite the improvement obtained by the aforementioned attention mechanism, the following question remains open: "Is there a better way to obtain a representation for the frequency characteristics of the input image?"

In this paper, a CNN architecture with an efficient and plug-and-play Sub-band Pyramid Attention (SPA) is proposed as an alternative for the existing channel attention models. Based on the wavelet decomposition, the SPA module performs a more fine-grained frequency selection that weighs the sub-bands at different levels. The SPA module exhibits a restoration performance while preserving the detailed textures with a negligible increase in computational complexity. Experimental results on real image denoising confirm the superiority of the proposed method. We also review how the utilized pyramid level influences the denoising performance and the generalization capability of the SPA blocks. The proposed SPA method with a proven superior performance can replace the existing attention mechanism and is easily applicable to DL networks with arbitrary structures. Our study highlights how leveraging fundamental knowledge in image processing can improve the performance of DL methods.
\begin{figure}[]
\begin{center}
\centerline{\includegraphics[width=0.99\columnwidth]{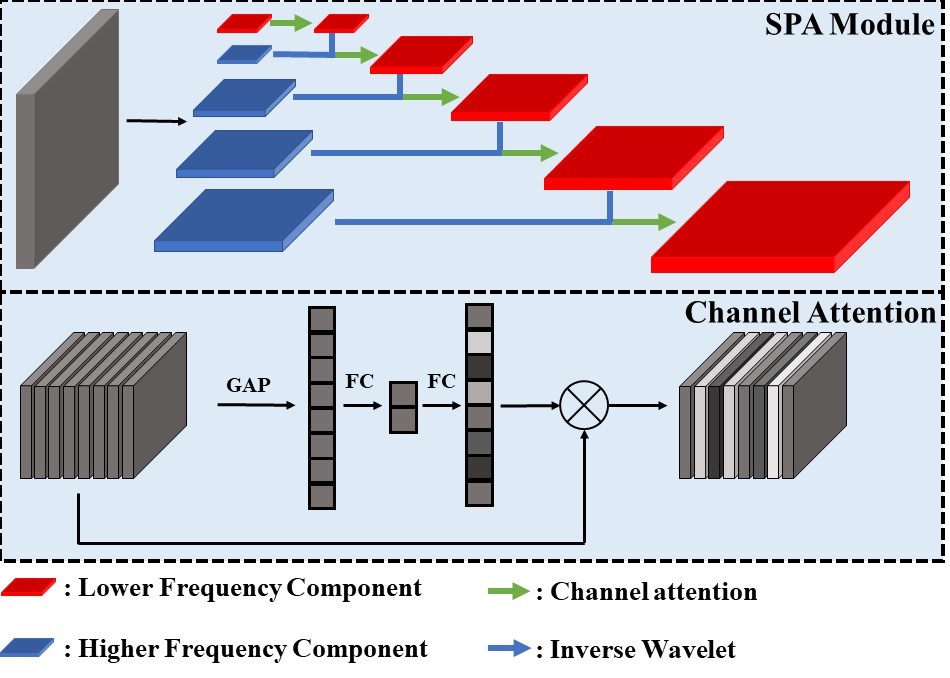}}
\caption{The sub-band pyramid attention for frequency components selection.}
\label{fig:spa_module}
\end{center}
\end{figure}

\begin{figure*}[]
\begin{center}
\centerline{\includegraphics[width=1.6\columnwidth]{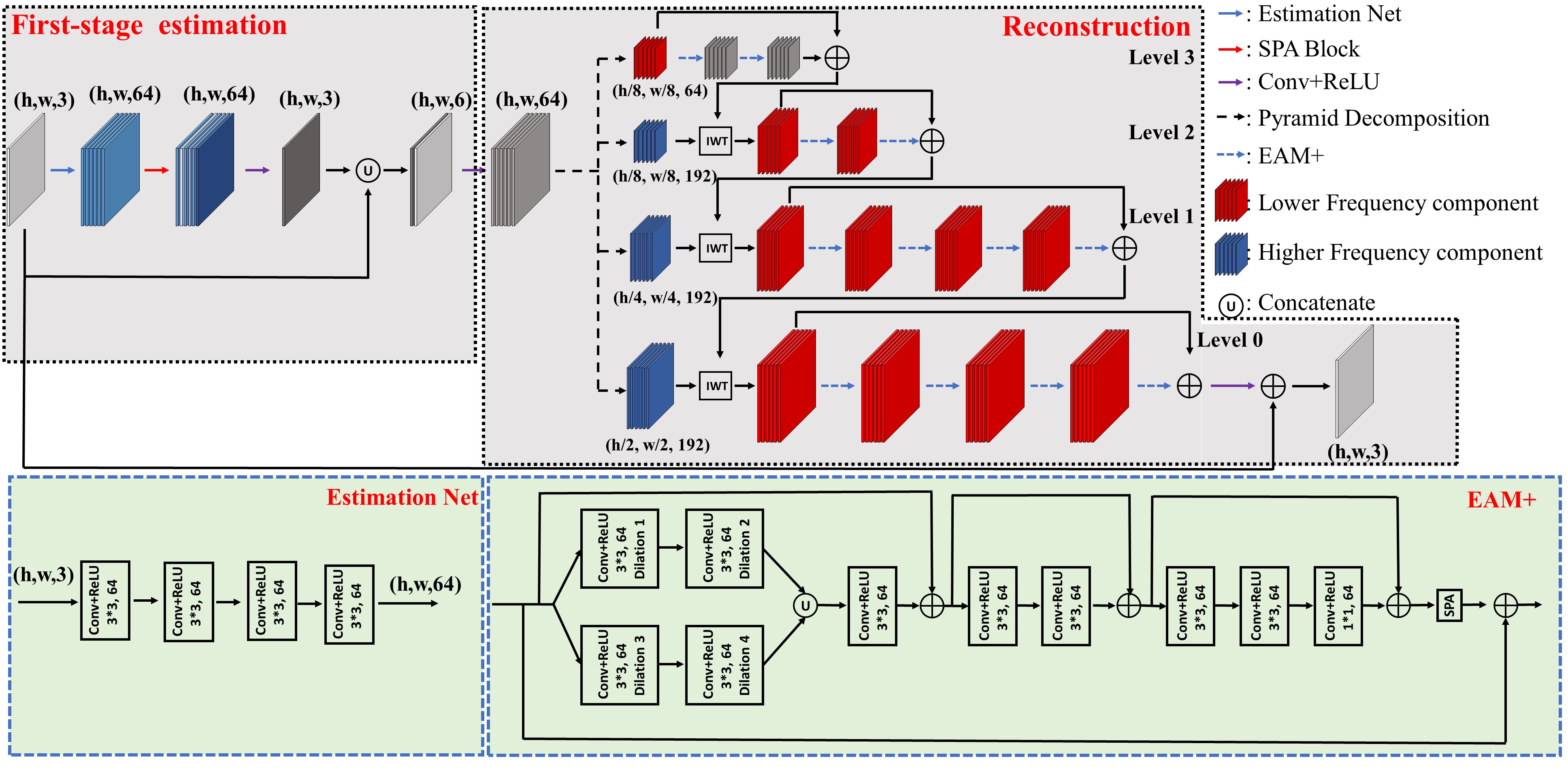}}
\caption{The illustration of the entire network architectures. The network is divided into noise estimation and reconstruction stage. The model in noise estimation stage is built by a plaint CNN with SPA blocks. The network in reconstruction is based on sub-band pyramid. The core component of the reconstruction network is a modified EAM~\cite{anwar2019real}, that we replaced the naive channel attention with SPA, and named it as EAM+.}
\label{fig:network}
\end{center}
\end{figure*}

\section{Method}

\begin{figure}[]
\begin{center}
\centerline{\includegraphics[width=0.8\columnwidth]{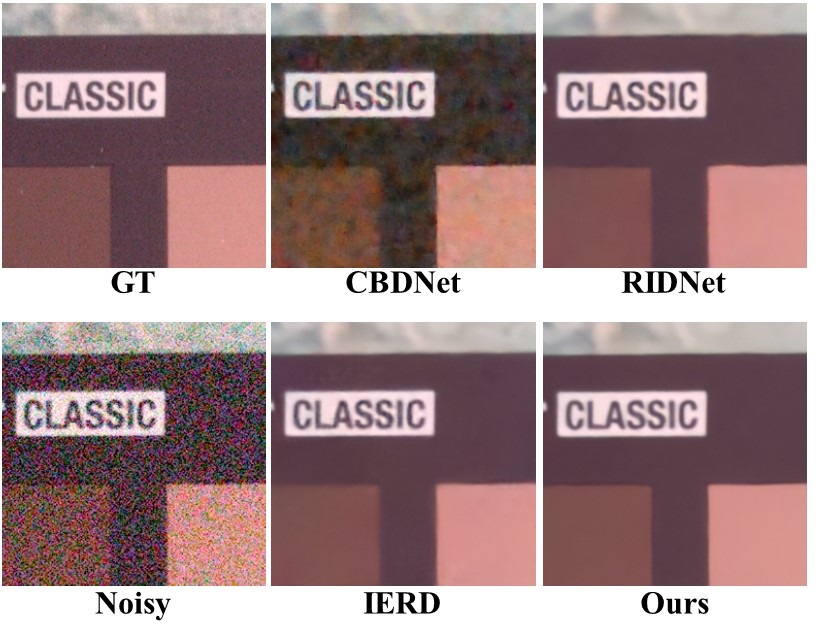}}
\caption{A challenging example from SIDD dataset~\cite{abdelhamed2018high}. Our model performs a better color and edge preservation property. Please zoom in for better vision.}
\label{fig:sidd}
\end{center}
\end{figure}

\begin{figure}[]
\begin{center}
\centerline{\includegraphics[width=0.8\columnwidth]{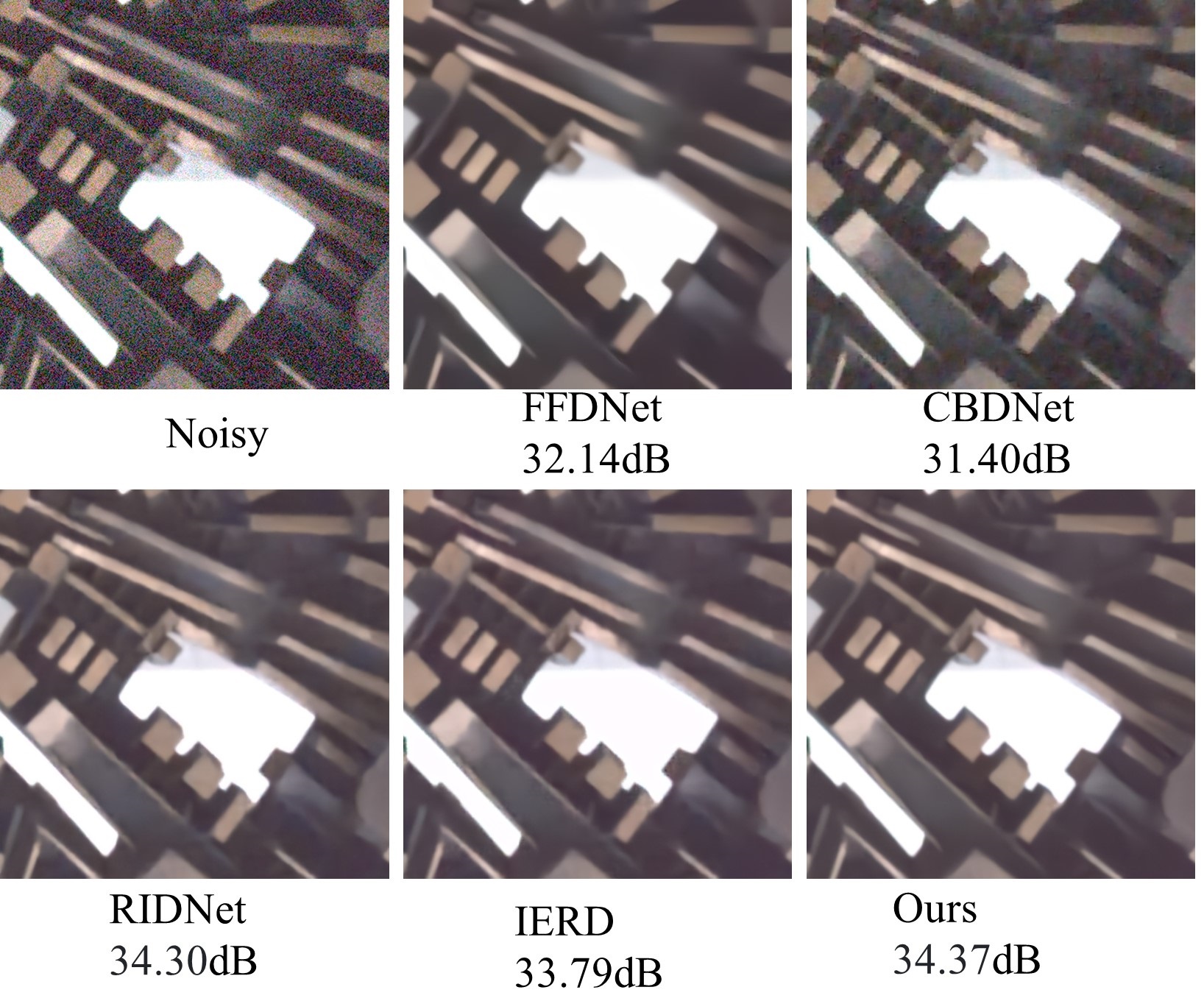}}
\caption{Comparison of our method against some popular image denoising methods on DnD dataset. Details can be better viewed by zooming the page.}
\label{fig:dnd}
\end{center}
\end{figure}

\subsection{Frequency Sub-band Pyramid}

The 2-D discrete wavelet transform (DWT) is a powerful tool for analyzing image structures. DWT decomposes the image into four sub-bands using four orthogonal convolutional filters, including one low pass filter $f_{LL}$, and three high pass filters $f_{LH}$, $f_{HL}$, and $f_{HH}$. Haar wavelet is a popular mother wavelet, which is also used in SPA, includes four orthogonal filters defined as: $f_{LL}$ = $ \left[ \begin{smallmatrix} +1 & +1 \\ +1 & +1 \end{smallmatrix}  \right] $, $f_{LH}$ = $ \left[ \begin{smallmatrix} -1 & -1 \\ +1 & +1 \end{smallmatrix}  \right]$, $f_{HL}$ = $ \left[ \begin{smallmatrix} -1 & +1 \\ -1 & +1 \end{smallmatrix}  \right] $, and $f_{HH}$ = $ \left[ \begin{smallmatrix} +1 & -1 \\ -1 & +1 \end{smallmatrix}  \right] $. Hence, four sub-bands are generated by convolving the input image $x$ with these filters to obtain: $x_{LL}=(f_{LL}\otimes x)$, $x_{LH}=(f_{LH}\otimes x)$, $x_{HL}=(f_{HL}\otimes x)$, and $x_{HH}=(f_{HH}\otimes x)$, where $\otimes$ is the convolution operator. Moreover, the biorthogonal property of DWT enables an easy and lossless reconstruction of the original image using the inverse transformation of Haar wavelet $IWT(x_{LL},x_{LH},x_{HL},x_{HH})$ as follows:
\begin{scriptsize}
\begin{flalign}
\begin{split}
&x(2i-1,2j-1) =\frac{(x_{LL}(i,j)-x_{LH}(i,j)-x_{HL}(i,j)+x_{HH}(i,j))}{4},\\
&x(2i-1,2j) =\frac{(x_{LL}(i,j)-x_{LH}(i,j)+x_{HL}(i,j)-x_{HH}(i,j))}{4},\\
&x(2i,2j-1) =\frac{(x_{LL}(i,j)+x_{LH}(i,j)-x_{HL}(i,j)-x_{HH}(i,j))}{4},\\
&x(2i,2j) =\frac{(x_{LL}(i,j)+x_{LH}(i,j)+x_{HL}(i,j)+x_{HH}(i,j))}{4}.
\end{split}
\end{flalign}
\end{scriptsize}

In this work, we propose to build a Frequency Sub-band Pyramid using wavelet decomposition. A Frequency Sub-band Pyramid consists of multi-level frequency components of an image or feature maps. Given an input $x_{0}$, DWT decomposes it into a set of high frequency components $X_{1H}=[x_{1HH},x_{1HL},x_{1LH}]$ and low frequency component $x_{1LL}$. After further decomposition using multi-level DWT for $n$ iterations, one low frequency component $x_{nL}$ and $n$ sets of high frequency components $X_{1H}, ... , X_{nH}$. Stacking these components from first to the last level forms a sub-band pyramid that repents the low-to-high frequency properties of the features to be modeled.




\subsection{Sub-band Pyramid Attention}
A channel attention~\cite{hu2018squeeze} module is typically formulated as: 
\begin{equation}
    x^{'}=x*\sigma (f_2(ReLU(f_1(G_x))))
\end{equation}
for a 3D input $x\in C\times H \times W$, a global descriptor $G_x \in C\times 1 \times 1$, which represents the statistics of each input map generated by the GAP. The functions $f_1$ and $f_2$ refer to two fully connected layers activated by the Rectified Linear Units ($ReLU$)~\cite{nair2010rectified} and sigmoid function ($\sigma$). The channel attention modules capture the channel dependencies from the global descriptor of the entire input, which is too coarse and may lead to information loss. The presented SPA exploits a more fine-grained channel-wise correlation with a new strategy based on naive channel attention. Overall, the decomposition results $[x_nL,x_nH, ..., x_2H, x_1H]$ of an input $x$ obtained by the Frequency Sub-band Pyramid are re-calibrated by the channel attention from lower to higher frequency levels as shown in Fig.\ref{fig:spa_module}. After being processed by the channel attention module, each lower frequency component was concatenated with its corresponding higher frequency component. The inverse wavelet transform (IWT) is used to build the lower-frequency components layer by layer, starting from the top layer to the base layer until the entire feature map with the original size (i.e., size of the input image) is reconstructed. The SPA module performs a more precise frequency selection mechanism than the naive channel attention approach by this operation. The SPA explicitly calibrates the dependencies between the feature channels while selecting the desired frequency component inside each feature map.

\subsection{Network Overview}
\label{sec:network}
The network used in this work comprises two stages where the first stage performs the noise estimation, and the second stage performs the reconstruction, as shown in Fig.\ref{fig:network}. For a noisy input $x\in C,H,W$, the first-stage $F_{e}$ can be regarded as a pixel-wise noise level estimation $x^{'}=F_{e}(x)$, where $x^{'}\in C,H,W$ is the estimation of the noisy maps for each input channel. The first-stage estimation includes four $64$-channel convolutional layers followed by a ReLU activation, a SPA block, and a $3$-channel convolutional layer. The filter size is $3\times3$ for each convolutional layer in the first-stage estimation. The SPA block is considered a frequency estimator to selectively suppress the redundant information from the extracted features. The estimation results are stacked with the input along the channels as $[x,x^{'}]$ and are fed to the second stage. 

Similar to the SPA block, the reconstruction stage is also designed based on the wavelet pyramid. This network consists of two convolutional layers (the first and last layers) and four sub-networks. The first convolutional layer extracts shallow features from the input image and the estimated noisy map. A level-$3$ wavelet pyramid of the feature is built and is further processed by the sub-networks. The level-$3$ to level-$1$ sub-networks process the low-pass sub-bands $x_{3LL},x_{2LL},x_{1LL}$, and the level-$0$ sub-networks processes the basis features $x_{0LL}$. The sub-networks are built based on enhancement attention modules (EAM) proposed in RIDNet~\cite{anwar2019real}. We replace the channel attention blocks in EAM with the proposed SPA blocks and further named the modules as EAM+. Each of the level-$3$ and level-$2$ sub-networks consists of two EAM+, and each of the level-$1$ and level-$0$ sub-networks consists of four EAM+. The sub-networks operate in a top-down manner, where each sub-network receives the lower-frequency map $x_{iLL}$ of the wavelet's current layer as its input and extends it to the entire map of this current layer, which is equivalent to the low-frequency map of the wavelet's previous layer. This information is passed to the next subnetwork until the full-size map is recovered. 



\section{Experiments}
\subsection{Setups}
We use the Smartphone Image Denoising Dataset (SIDD)~\cite{abdelhamed2018high} and  Darmstadt Noise Dataset (DnD)~\cite{plotz2017benchmarking} datasets for image denoising. The SIDD dataset provides 320 clean and noisy image pairs for training along with 1280 image pairs for validation. DnD dataset contains 50 pairs of real-world noisy and noise-free scenes. It provides bounding boxes with size $512\times512$ of 1000 Regions of Interests (ROIs) for 50 scenes for generating testing data. 

We implemented our model using the Pytorch Framework~\cite{paszke2017automatic} and trained it with a Tesla P40 GPU. We equipped the network with level-3 SPA blocks. The hyper-parameters of the network is defined in Fig.\ref{fig:network}. We used $512\times512$ patches cropped from the SIDD~\cite{abdelhamed2018high} training set to train our model and used the DnD~\cite{plotz2017benchmarking} and validation set of the SIDD dataset to evaluate the reconstruction performance. Data augmentation by applying random rotations at $90$, $180$, and $270$ degrees and horizontal flipping was used in the training phase. Peak Signal-to-Noise Ratio (PSNR) is used as the evaluating metric, while Mean Absolute Error (MAE) is used as the loss function. The model is trained by the Adam optimizer~\cite{kingma2014adam} with an initial learning rate $1e-4$. We trained the model at $2.5e5$ iterations and halved the learning rate for each of the $1e5$ iterations. 

\begin{table}[]
\begin{center}
\begin{tabular}{|llll|}
\hline
Method                       & Blind/Non-blind                       &PSNR                         &SSIM  \\ \hline
\multicolumn{1}{|l|}{CDnCNN-B~\cite{zhang2017beyond}} & Blind  &  \multicolumn{1}{|l}{32.43} & \multicolumn{1}{l|}{0.7900}        \\ 
\multicolumn{1}{|l|}{TNRD~\cite{chen2016trainable}} & Non-blind  & \multicolumn{1}{|l}{33.65} & \multicolumn{1}{l|}{0.8306}        \\ 
\multicolumn{1}{|l|}{LP~\cite{burger2012image}} & Non-blind  & \multicolumn{1}{|l}{34.23} & \multicolumn{1}{l|}{0.8331}        \\ 
\multicolumn{1}{|l|}{FFDNet~\cite{zhang2018ffdnet}} & Non-blind & \multicolumn{1}{|l}{34.40} & \multicolumn{1}{l|}{0.8474}        \\ 
\multicolumn{1}{|l|}{BM3D~\cite{dabov2007image}} & Non-blind   & \multicolumn{1}{|l}{34.51} & \multicolumn{1}{l|}{0.8507}       \\ 
\multicolumn{1}{|l|}{WNNM~\cite{gu2014weighted}} & Non-blind  & \multicolumn{1}{|l}{34.67} & \multicolumn{1}{l|}{0.8646}        \\ 
\multicolumn{1}{|l|}{KSVD~\cite{aharon2006k}} & Non-blind   & \multicolumn{1}{|l}{36.49} & \multicolumn{1}{l|}{0.8978}       \\ 
\multicolumn{1}{|l|}{MCWNNM~\cite{xu2017multi}}  & Non-blind & \multicolumn{1}{|l}{37.38} & \multicolumn{1}{l|}{0.9294}        \\ 
\multicolumn{1}{|l|}{FFDNet+~\cite{zhang2018ffdnet}} & Non-blind & \multicolumn{1}{|l}{37.61} & \multicolumn{1}{l|}{0.9415}       \\ 
\multicolumn{1}{|l|}{TWSC~\cite{xu2018trilateral}}  & Non-blind  & \multicolumn{1}{|l}{37.96} & \multicolumn{1}{l|}{0.9416}       \\ 
\multicolumn{1}{|l|}{CBDNet~\cite{guo2019toward}} & Blind & \multicolumn{1}{|l}{38.06} & \multicolumn{1}{l|}{0.9421}           \\ 
\multicolumn{1}{|l|}{RIDNet~\cite{anwar2019real}} & Blind & \multicolumn{1}{|l}{39.25} & \multicolumn{1}{l|}{0.9528}           \\ 
\multicolumn{1}{|l|}{IERD~\cite{anwar2020identity}}   & Blind  & \multicolumn{1}{|l}{39.30} & \multicolumn{1}{l|}{0.9531}          \\ \hline
\multicolumn{1}{|l|}{Ours}   & Blind & \multicolumn{1}{|l}{\textbf{39.48}} & \multicolumn{1}{l|}{\textbf{0.9580}}           \\ \hline
\end{tabular}
\caption{PSNR and SSIM of the denoising methods evaluated on DnD~\cite{plotz2017benchmarking} dataset.}
\label{table:DnD}
\end{center}
\end{table}

\subsection{Real Image Denoising}
Table.\ref{table:SIDD} represents the performances of the proposed method in term of image denoising and reconstruction in comparison with several benchmark methods using the SIDD validation set. It can be seen that our method obtains remarkable results and outperforms the most commonly used DL-based denoising algorithms. Fig.\ref{fig:sidd} presents an illustrative example which shows that our method holds a competitive color and edge preservation property. Also, the results for the DnD dataset are summarized in Table.\ref{table:DnD}. We compare the PSNR and SSIM~\cite{wang2004image} through the online evaluation system provided by the DnD~\cite{plotz2017benchmarking} official website. The results show that the proposed method achieves a 3.97 dB gain with respect to the best traditional algorithm, BM3D~\cite{dabov2007image}, while realizing a 1.87 to 0.18 dB gain over the DL-based methods~\cite{zhang2018ffdnet,guo2019toward,anwar2019real,anwar2020identity}. Fig.\ref{fig:dnd} shows the proposed method in restoring a noise-free image without over smoothing the details. 

\begin{table}[]
\begin{center}
\begin{tabular}{|l|l|l|l|l|l|}
\hline
\multicolumn{6}{|c|}{Method}                     \\ \hline
BM3D & FFDNet & CBDNet & RIDNet & IERD  & Ours  \\ \hline
30.88 & 29.20  & 30.78  & 38.71  & 38.82 & \textbf{39.55} \\ \hline
\end{tabular}
\caption{The quantitative results (PSNR) for the SIDD dataset~\cite{abdelhamed2018high}.}
\label{table:SIDD}
\end{center}
\end{table}


\subsection{Ablation Study} 
The major contribution of this work is proposing a network architecture based on SPA blocks, a novel implementation of attention mechanism SPA block. Although SPA blocks have already experienced remarkable improvements, our approach offers a new perspective to this problem and opens up new questions like "how the pyramid level influences the denoising performance of the SPA blocks" and "if the SPA blocks can boost the performance of other networks with arbitrary architectures." Our results prove the superiority of the proposed SPA block compared to the naive channel attention block. At the same time, it also justifies the effect of different levels of SPA block. We also conducted an ablation study on the proposed network architecture. We test PSNR on the SIDD validation set for levels-$0$ to level-$4$. Notably, we define the naive channel attention as level-$0$ pyramid (which means no pyramid). The second row of Table.\ref{table:ablation} shows that the higher pyramid level of the SPA blocks leads to a high PSNR and a better denoising performance. For evaluating the generalization of the SPA blocks, we conducted another ablation study by deploying SPA blocks at different levels in the RIDNet network as an alternative for its channel attention blocks. The third row Table.\ref{table:ablation} shows the results of RIDNet when equipped with SPA blocks of different pyramid levels. The results show that the SPA blocks generalize well in RIDNet while expecting to confirm that pyramids with more layers lead to better results. 

\begin{table}[]
\begin{center}
\begin{tabular}{|l|l|l|l|l|l|l|}
\hline
Methods & Level & 0     & 1     & 2     & 3     & 4     \\ \hline
Ours    & PSNR  & 39.24 & 39.33 & 39.47 & 39.55 & 39.57 \\ \hline
RIDNet  & PSNR  & 38.71 & 38.87 & 38.99 & 39.04 & 39.05 \\ \hline
\end{tabular}
\caption{Investigation of effects of the pyramid level of SPA block and generalization on RIDNet.}
\label{table:ablation}
\end{center}
\end{table}

\section{Conclusion}
A novel channel attention module called Sub-band Pyramid Attention (SPA) is proposed in this work. The SPA blocks are built upon wavelet decomposition to realize a joint sub-band channel attention. The SPA is implemented as a plug-and-play module, hence can be used for alternating the naive channel attention. The SPA block performs a more precise feature re-calibration that both re-scales the feature channels and the multi-level frequency components. In addition, a CNN architecture based on the wavelet pyramid was designed. The generalization of the SPA blocks was proven by the ablation study, which suggests that the SPA module is compatible with other network architectures and can be widely used in other networks to boost the image restoration performance.

\bibliography{main}
\bibliographystyle{IEEEtran}
\vspace{12pt}
\end{document}